\documentclass[12pt,oneside]{article}
\usepackage{epsfig, cite}
\usepackage{color}
\usepackage{ifthen}
\usepackage{graphicx}

\def\d{\Lambda}
\def\td{{\tilde \lambda}}

\def\ap{\alpha'}
\def\half{{1\over 2}}
\def\({\left(}
\def\){\right)}
\def\[{\left[}
\def\]{\right]}

\def\e{\begin{equation}}
\def\q{\end{equation}}
\def\m{\begin{eqnarray}}
\def\n{\end{eqnarray}}

\begin{document}
\thispagestyle{empty} \setcounter{page}{0}

\vspace{2cm}

\begin{center}
{\huge Observational Consequences of Quantum Cosmology}

\vspace{1.4cm}

Qing-Guo Huang

\vspace{.2cm}

{\em Interdisciplinary Center of Theoretical Studies, Chinese Academia Sinica,}\\
{\em P. O. Box 2735, Beijing 100080, China} \\
{\em and} \\
{\em Institute of Theoretical Physics, Chinese Academia
Sincia} \\
{\em P. O. Box 2735, Beijing 100080, China}

\end{center}

\vspace{-.1cm}

\centerline{{\tt huangqg@kias.re.kr}} \vspace{1cm}
\centerline{ABSTRACT}
\begin{quote}
\vspace{.5cm}

Our universe is born of a tunnelling from nothing in quantum
cosmology. Nothing here can be interpreted as a state with zero
entropy with the viewpoint of a dual CFT proposed by Verlinde. As a
reliable modification of the Hartle-Hawking wave function of the
universe, the improved Hartle-Hawking wave function proposed by
Firouzjahi, Sarangi and Tye gives many interesting observational
consequences which we explore in this paper. Fruitful observations
are obtained for chaotic inflation, including a detectable spatial
curvature and a negligible tunnelling probability for eternal
chaotic inflation. And we find that the tensor-scalar ratio and the
spatial curvature for brane inflation type models should be
neglected.

\end{quote}
\baselineskip18pt

\noindent

\vspace{5mm}

\newpage

In the last few years, the cosmological observations \cite{wmap,ex}
are successful in making the history of our universe clearer. Recent
expansion of our universe is accelerating, due to the dark energy
with negative pressure. As a simplest candidate for the dark energy,
the cosmological constant can fit today's data very well. Hot Big
Bang must happen in the early universe. In order to solve the
flatness, horizon and monopole puzzle etc. in Hot Big Bang model, a
procession with the expansion of the universe accelerating, named
inflation, should happen before the Hot Big Bang. Inflationary
models also naturally offer the seeds for the formation of the large
scale structure of our universe and the temperature fluctuations in
CMB.

However the initial condition for our universe is still unknown. Why
and how inflation happens before Hot Big Bang are still mysterious
in modern physics. Until now, the most attractive idea about the
origin of our universe is the Hartle-Hawking no-boundary wave
function $\Psi_{HH}$ of the universe \cite{hh} which says that our
universe is born of a tunnelling from nothing (here nothing means a
state without any classical spacetime \cite{av}). A number of other
conjectures on the origin of our universe are also discussed in
\cite{aas}. In general, we believe the quantum theory of gravity can
give us more insights on the origin of the universe. Unfortunately,
we still don't know the full quantum theory of gravity. String
theory is the only self-consistent candidate for quantum gravity.
Recently some advancements took place in string theory, notably
string compactifications \cite{sc}, which provide many stable or
meta-stable vacua with lifetimes comparable to or larger than the
age of our universe. The number of such vacua may be ${\cal{O}}
(10^{100})$ or larger. On the other hand, in the inflationary
scenario, our observed universe is a tiny patch of the whole
spacetime in the early universe and the whole space-time may contain
many disconnected regions which will grow up to many subuniverses
with different physical parameters. How to select a particular
vacuum where we live is still a big open question. A line to go
beyond the Anthropic principle is to reliably calculate the
tunnelling probability from nothing and pick out the vacuum with the
largest tunnelling probability (see, for example, \cite{avp}).

Hartle-Hawking wave function of the universe favors a universe with
cosmological constant $\d=0$. Inflation can not happen in such a
universe. It does not agree with the history of our universe. In
this short note, we prefer an improved Hartle-Hawking wave function
proposed by Firouzjahi, Sarangi and Tye in \cite{fst,st}. This wave
function offers a natural realization of the spontaneous creation of
inflationary universe from nothing. According to the Cardy-Verlinde
formula \cite{ev}, the evolution of a closed universe can be
described by a dual CFT and we can use it to make the meaning of
nothing clearer, i.e. nothing means that the initial state of our
universe corresponds to a state of CFT with zero entropy. We also
use this improved wave function of the universe to investigate
several typical inflationary models, such as chaotic inflation and
brane inflation type model. Many interesting observational
consequences are obtained.

Let us start with the action for Einstein theory in four dimensional
spacetime with a positive cosmological constant $\d$, \e \label{act}
S_E=\int d^4x \sqrt{-g} \left(-{R \over 16 \pi} + \d \right). \q We
work on the unit with Newton coupling constant $G_N=1$ in the whole
note. The Euclidean metric is given by \e \label{eum}
ds^2=\sigma^2\left[ d\tau^2+a^2(\tau)d\Omega_3^2 \right], \q where
$\sigma^2=2/3\pi$ and $d\Omega_3^2$ is the metric on a unit $S^3$.
Since the volume of the spatial flat or open universe is infinite,
\footnote{If the topology or the configuration of the universe is
nontrivial, a flat or open universe is also possibly tunnelling from
nothing, see for instance \cite{fl,bm}. } the tunneling probability
from nothing to any one of them is suppressed and it is only
possible that a closed de Sitter universe emerges. The equation of
motion for the universe is given by \e \label{ea}\left({\dot a \over
a} \right)^2+{1 \over a^2}={16\d \over 9}. \q The tunneling point
corresponds to $\dot a/a=H=0$ and our universe started with a finite
size, $a=3/(4\sqrt{\d})$, not $a=0$, which says that Hartle-Hawking
wave function can help us to avoid the cosmic singularity.

Recalling the proposal by Verlinde in \cite{ev}, the entropy of a
$(1+1)$-dimensional CFT is given by Cardy formula \e \label{cf}
S=2\pi \sqrt{{c \over 6} \left(L_0-{c \over 24} \right)}, \q which
is universally valid in arbitrary dimensional spacetime if we define
the central charge $c$ in terms of the Casimir energy. Via some
identifications, for instance, the entropy of this CFT with the
$(n-1){HV \over 4G}$(see \cite{ev} in details), where $H$ is the
Hubble constant and $V$ is the volume of $(n+1)$-dimensional closed
Friedman-Robertson-Walker universe, the Cardy formula (\ref{cf})
exactly turns into the $(n+1)$-dimensional Friedman equation for a
closed universe. Or the entropy of a closed universe with the
viewpoint of a dual CFT can be given by \e \label{enu}S_H = (n-1){HV
\over 4G_N}. \q When the universe was born, the expansion rate of
the universe $H=\dot a /a$ equals zero. Using eq. (\ref{enu}), the
entropy of the universe equals zero as well. Thus the meaning of
nothing in quantum cosmology can be interpreted as a state with
entropy being zero.

According to Hartle-Hawking wave function, the tunnelling
probability from nothing to a closed de Sitter universe with
cosmological constant $\d$ is \e \label{hhp} P_{HH}=|\Psi_{HH}
|^2\simeq \exp{\left({3 \over 8 \d} \right)}. \q This wave function
is not normalizable and picks out $\d=0$ with the largest tunnelling
probability. It contradicts with the history of our universe. The
size of the universe is roughly $1/\sqrt{\d}$, a macroscopic size.
Tunnelling is a pure quantum phenomenon and the tunnelling
probability is suppressed for a quantum system with macroscopic
size. Naturally we expect that the probability for tunnelling to a
universe with $\d=0$ is suppressed by the effects of decoherence and
a universe with microscopic size naturally emerges. This is just the
idea proposed by Firouzjahi, Sarangi and Tye to improve
Hartle-Hawking wave function \cite{fst,st}. The decoherence comes
from the fluctuations of the metric. After integrating out these
fluctuations, a new term which behaves just like ordinary radiation
appears in the effective Lorentzian action \e \label{nact} S \simeq
\half \int d \tau \left(-a{\dot a}^2+a-\alpha
 a^3-{\nu \over \alpha^2 a} \right), \q where $\alpha=16\d/9$ and $\nu$
is a constant which measures the number of perturbative modes and
depends on the UV cutoff for the gravitational interaction in
quantum theory of gravity. Roughly the value of $\nu$ relates to
this UV cutoff $M$ by (see \cite{st}) \e \label{mnu}\nu=M^4/(72
\pi^6). \q In string theory, this UV cutoff is naturally taken as
string scale $M_s=1/\sqrt{\ap}$, a high energy scale. Since we do
not know string scale exactly, we can not determine the value of
$\nu$. In particular, in the framework of KKLT \cite{sc}, the
effective string scale in different throats can take different
values, which depends on the value of warp factor. It is natural
that the UV cutoff in eq. (\ref{mnu}) is lower than the Planck
energy scale $M_{pl}$ and then $\nu \leq 10^{-5}$. Here we will fix
it by using the observational data.

Now the equations of motion following from the improved action
(\ref{nact}) are given by \m \({\dot a \over a} \)^2+{1 \over
a^2}&=&\alpha+{\nu \over \alpha^2 a^4}, \\ {\ddot a \over
a}&=&\alpha-{\nu \over \alpha^2 a^4}. \label{mem}\n The initial size
of the universe is \e \label{ins} a={1 \over
\sqrt{2\alpha}}\left(1+\sqrt{1-{4\nu \over \alpha}} \right)^{1/2}.
\q Here we require $\alpha \geq 4\nu$, which is just the condition
for the existence of the instanton solution. The reason we choose
this solution (\ref{ins}) is that the right hand side of eq.
(\ref{mem}) is positive and the expansion rate of the universe can
speed up from zero after the universe was born. The author of
\cite{fst,st} also calculate the tunnelling probability according to
this improved acton as \e \label{tp} P \simeq \exp {\cal{F}}= \exp
\left({3 \over 8\d}-{27 \nu \over 32 \d^2} \right). \q We need to
remind that the instanton solution is destroyed and the tunnelling
probability equals zero when $\d<9\nu/4$. Since the UV cutoff is a
high energy scale, the effective cosmological $\d$ is large and
inflation can naturally occur.

In the following, we will investigate the consequences for several
typical inflation models by using this improved tunnelling
probability (\ref{tp}). Here we make the loose-shoe approximation
\cite{kt}, $\phi=\hbox{const}$ and the potential $V(\phi)$ term is
equivalent to a cosmological constant. Therefore eq. (\ref{tp}) is
recovered.

First we work on the inflation model with only one fine tuning
parameter, for example, chaotic inflation model \cite{al}. The
potential for the inflaton field $\phi$ is given by \e
\label{pci}V(\phi)=\lambda \phi^n, \q where $n \geq 1$. The
parameter needed to be fine tuned is the coupling constant $\lambda$
which can be fixed by the amplitude of the power spectrum. Or
equivalently, there is no free parameter at all if we take the
amplitude of the power spectrum as an input. The equations of motion
in the slow rolling limit can be written as \m \label{cem} H^2
&\simeq& {8\pi V(\phi) \over 3}={8\pi \over 3} \lambda \phi^n, \\
3H{\dot \phi} &\simeq& -V'(\phi)=-n\lambda \phi^{n-1}. \n The number
of e-folds before the end of inflation is \e N=\int H dt=\int {H
\over \dot \phi} d \phi \simeq {4 \pi \over n} \phi_N^2. \q Or
equivalently, the value of $\phi$, namely $\phi_N$ at the number of
e-folds $N$ before the end of the inflation is \e \phi_N=\sqrt{{n
\over 4\pi}N}. \q When $\phi=\phi_N$, the potential becomes \e
V(\phi_N)=\td N^{n\over 2}, \q with \e \td=\lambda \left({n \over
4\pi} \right)^{n\over 2}. \q The tunnelling probability (\ref{tp})
can be obtained as \e \label{ctp} P \simeq \exp{\cal{F}} = \exp
\left({3 \over 8 \td N^{n/2}}-{27\nu \over 32 \td^2 N^n} \right). \q
Maximizing ${\cal{F}}$ in eq. (\ref{ctp}), we obtain \e
\label{cnn}N=N_m=\left({9\nu \over 2 \td} \right)^{2/n}. \q Now the
tunnelling probability is given by \e \label{ccp} P \simeq
\exp{\cal{F}} = \exp \left({1 \over 24 \nu} \right)=\exp \left({3
\over 16 \td N_m^{n/2}} \right). \q

The amplitude of the power spectrum $\delta_H$ is \e
\label{cps}\delta_H={1 \over 5\pi} \left|{H^2 \over \dot \phi}
\right| \simeq \left({2^7 \over n\cdot 3 \cdot 5^2} \right)^\half
\td^\half N^{n+2 \over 4}, \q where $N$ evaluates the number of
e-folds when the fluctuations just stretched outside the Hubble
horizon during the period of inflation. The COBE normalization is
$\delta_H\simeq 2\times 10^{-5}$ for $N \sim 60$. If $n=4$, we find
$\td \simeq 4.3\times 10^{-15}$ and ${\cal F}_{N_m} \simeq 4.4
\times 10^{13}/N_m^2$. For a roughly spatial flat universe, the
total number of e-folds should not be smaller than 60. Using eq.
(\ref{cnn}), we can get a low bound on the parameter $\nu$ as $\nu
\geq 3.4\times 10^{-12}$ and an upper bound on the tunnelling
probability with ${\cal F} \leq {\cal F}_{60} = 1.2 \times 10^{10}$.
Combining with eq. (\ref{mnu}), we can get a low bound on the UV
cutoff in eq. (\ref{mnu}) as $M \geq 0.02 M_{pl}$, where
$M_{pl}=G_N^{-1/2}=1.2\times 10^{19}$Gev is the Planck energy scale
in four dimensions. In string theory, we take this UV cutoff as
string scale, which means that the string energy scale should not be
lower than $0.02 M_{pl}$. Here we need to keep in mind that a
successful realization of a chaotic inflation in string theory is
still not known.

For a given $\nu$, the distribution for the number of e-folds can be
written as \e \label{cgnn} P \simeq \exp \left({3 \over 8 \td}
\left({1 \over N^{n/2}}-{N_m^{n/2} \over 2 N^n} \right) \right) \sim
\exp \left(-{(N-N_m)^2 \over 2\sigma^2} \right), \q which is a
Gaussian distribution with variance \e \label{sm} \sigma=\sqrt{32
\td \over 3n^2} N_m^{1+{n \over 4}}. \q Here we drop out a whole
constant factor in the last step and take a limit with $|N-N_m| \ll
N_m$ in eq. (\ref{cgnn}). For $n=4$ and $N_m =60$, the variance
roughly equals $2\times 10^{-4}$. Therefore, for a given $\nu$, the
number of e-folds must be $N_m$ given by eq. (\ref{cnn}) at a high
level of statistical significance.

As a generic phenomenon, eternal inflation is common to a very wide
class of inflation models \cite{vl}. During the period of inflation,
the evolution of the inflaton field $\phi$ is influenced by quantum
fluctuations, which can be pictured as a random walk of the field
with a step $\delta \phi \sim H/2\pi$ on a horizon scale (Hubble
scale $H^{-1}$) per Hubble time $\Delta t \sim H^{-1}$. During the
same epoch, the variation of the classical homogeneous inflaton
field rolling down its potential is $\Delta \phi \sim |\dot \phi|
H^{-1}$. If the classical variation is smaller than the quantum
fluctuations, the role played by fluctuations becomes significant
and the inflaton field can walk up the potential, rather than roll
down the potential in some spacetime regions. If the eternal
inflation happened, the wave function of the universe can not help
us to select a particular vacuum where we live. The eternal
inflation provides a natural arena for the Anthropic principle.
Different parts of spacetime can be characterized by different
effective values of nature. The values we observe are determined by
Anthropic selection \cite{adl}.

We can estimate that the eternal chaotic inflation with potential
(\ref{pci}) happened when $\delta \phi \geq \Delta \phi$, or \e
\label{cetn}\phi \geq \phi_*=\left({3 n^2 \over 128 \pi
\lambda}\right)^{1 \over n+2}. \q If inflation started with
$\phi=\phi_*$, the total number of e-folds will be \e \label{ccetn}
N_*={4\pi \over n} \left({3 n^2 \over 128 \pi \lambda} \right)^{2
\over n+2}. \q For $n=4$, the number of e-folds for the chaotic
eternal inflation must be as large as $N_*\simeq 4.4 \times 10^4$.
According to eq. (\ref{ctp}), the tunnelling probability is
extremely suppressed with ${\cal F}_{N_*} \simeq 4 \times 10^4 \ll
{\cal F}_{60}$. The tunnelling probability for the chaotic inflation
starting with $\phi_*$ or larger is negligible. Therefore the
improved Hartle-Hawking wave function avoids the chaotic eternal
inflation.

In fact, there is a debate about the correct sign in the exponent in
the tunnelling probability from nothing to a closed de Sitter
universe. In \cite{alav}, the authors proposed that the tunnelling
probability is $\exp\left(-{3\over 8\Lambda}\right)$. It prefers a
large cosmological constant and the chaotic eternal inflation
naturally emerges. Recently the authors in \cite{sst} pointed out
that the decoherence effect enhances the tunnelling probability if
we use the results in \cite{alav}. It seems quite unreasonable. It
is possible that the improved Hartle-Hawking wave function can lead
to some observational results.

We also notice that the improved Hartle-Hawking wave function
(\ref{ccp}) prefers a universe with the total number of e-folds as
small as possible. This is a very subtle point. It means a closed
universe with large enough spatial curvature to be detected can
naturally emerge even for chaotic inflation. It is quite different
from the traditional point of view \cite{cl}. In fact, the WMAP
teams find that the best fit model to the first year data of WMAP
combined with other data sets is slightly closed with
$\Omega_k=-0.02\pm 0.02$ in \cite{wmap}, where
$\Omega_k=1-\Omega_m-\Omega_\d$. Other results are all consistent
with a slightly closed universe (for example, see \cite{teg}).
Additionally, the WMAP results \cite{wmap} confirm that the
amplitude of the quadrupole in the temperature power spectrum is low
compared with the predictions of $\d$CDM models seen by COBE
\cite{hin} as well. In \cite{ge}, Efstathiou proposed that the low
CMB quadrupole amplitude may be related to a truncation of the
primordial fluctuation spectrum on the curvature scale in a closed
universe. But it is very difficult to obtain a realistic model of a
closed inflationary universe before. Here we find that the improved
Hartle-Hawking wave function can give us a possible mechanism to
provide a detectable spatial curvature and help us to explain the
deficit of quadrupole in the temperature power spectrum for chaotic
inflation if the gravitational UV cutoff $M$ or string scale $M_s$
in string theory is roughly $0.02 M_{pl}$. \footnote{There are also
many other works on the deficit of the quadrupole, see \cite{enh}. }

The scalar power spectral index $n_s$ and the tensor-scalar ratio
$r$ is still not clear. Current measurements suggest that $n_s
=0.980 \pm 0.020$ with $68 \%$ confidence and $r \leq 0.36$ with $95
\%$ confidence by combining WMAP and the SDSS galaxy survey
\cite{us}. A fit using WMAP and BOOMERANG CMB data and the SDSS and
2dFGRS galaxy surveys \cite{cjm} gives $n_s=0.950\pm 0.020$ at $68
\%$ confidence level. The tensor-scalar ratio for the chaotic
inflation is given by $r=4n/N \sim {\cal{O}} (10^{-1})$ and the
spectral index is $n_s=1-{n+2 \over 2N}$ ($n_s\simeq 0.967$ for
$n=2$ and $n_s\simeq 0.95$ for $n=4$). Both the spectral index and
the tensor-scalar ratio for the chaotic inflation are close to the
observational bounds. We expect the cosmological observations can
provide stronger evidence to support or rule out this model in the
near future.

Similarly we also investigate the inflation models with two fine
tuning parameters, for example, brane inflation type model with
potential \e \label{bpt} V(\phi)=V_0 \left(1-{\mu^n \over \phi^n}
\right), \q where $n \geq 1$. There are two parameters, $V_0$ and
$\mu$. The evolution of inflation is dominated by $V_0$ and the
tunnelling probability from nothing is \e \label{btp} P \simeq \exp
{\cal F} = \exp \left({3 \over 8V_0}-{27 \nu \over 32 V_0^2}
\right). \q The tunnelling probability goes to its maximum value
when \e \label{bmv}V_0=V_{0m}={9\nu \over 2}, \q with \e \label{bmp}
P \simeq \exp {\cal F} =\exp \left({1 \over 24 \nu}\right) = \exp
\left({3 \over 16 V_0} \right). \q The tunnelling probability at
$V_0$ close to $V_{0m}$ is negligible compared to that at $V_{0m}$.
If the amplitude of the power spectrum is taken as an input, we
obtain a relationship between $V_0$ and $\mu$. Thus there is only
one free parameter. The equations of motion can be written down as
\m H^2 &\simeq& {8\pi V_0 \over 3}, \\ 3H{\dot \phi} &\simeq& -{n
V_0 \mu^n \over \phi^{n+1}}. \n The number of e-folds before the end
of inflation is \e \label{cne} N=\int H dt={8 \pi \over
n(n+2)}{\phi_N^{n+2} \over \mu^n}. \q Or equivalently, the value of
$\phi$, namely $\phi_N$ at the number of e-folding $N$ before the
end of the inflation is given by \e \label{cpn} \phi_N=\left({n(n+2)
\over 8\pi}\mu^n N \right)^{1 \over n+2}.\q The amplitude of the
power spectrum can be expressed as \e \delta_H \simeq \left({2^9
\cdot \pi \over 3 \cdot 5^2 \cdot n^2} \right)^\half
\left({n(n+2)\over 8\pi} \right)^{n+1 \over n+2}{V_0^\half \over
\mu^{n \over n+2}} N^{n+1 \over n+2}. \q For $n=2$, using COBE
normalization, we obtain \e V_0=9\times 10^{-13} \mu, \label{bcvm}\q
and ${\cal F} \simeq 2.1 \times 10^{11}/\mu$. If $\nu$ is given, the
value of parameter $\mu$ is determined by using eq. (\ref{bmv}) and
(\ref{bcvm}). On the other hand, smaller the parameter $\mu$, or
equivalently lower the inflation energy scale, larger the tunnelling
probability. The amplitude of the tensor (gravitational wave)
fluctuations only depends on the inflation energy scale. According
to the improved Hartle-Hawking wave function, an inflation model
with small amplitude of the primordial tensor fluctuations is
favored. We can expect that the tensor-scalar ratio is too small to
be detected for this inflation model. We also notice that there is
no stringent constraint on the number of e-folds and then the
spatial curvature should be neglected. The spectral index for this
model is $n_s=1-{n+1 \over n+2}{2 \over N}$ ($n_s\simeq 0.975$ for
$n=2$ and $n_s\simeq 0.972$ for $n=4$), a red spectrum as well. The
similar results are also obtained for the inflation model with
potential $V(\phi)=V_0 (1-\phi^n / \mu^n)$, where $n>2$.

In summary, the improved Hartle-Hawking wave function offers many
interesting observational consequences. It naturally provides a
possible mechanism, otherwise hard for us to understand before, to
obtain a slightly closed universe with a detectable spatial
curvature for chaotic inflation, which is consistent with the
current results of the cosmological measurements. However the
observational consequences for the brane inflation type model are
very boring, no detectable spatial curvature and tensor
fluctuations, and more fine tuning parameters are included. We
expect that the cosmological observations will give us more
conclusive results in the near future.

An old but very important problem in Hot Big Bang model is where the
initial expansion rate of the universe came from. Quantum cosmology
give us a scenario for the universe with initial expansion rate
being zero. A positive cosmological constant provide the force to
push the expansion of the universe. When the positive cosmological
constant dominated the evolution of the universe, the expansion of
the universe is accelerating, namely inflating, and the expansion
rate is quite large. At the end of the inflation, Hot Big Bang
happened. This is roughly the history of the universe. However,
there is still one more question: where does the cosmological
constant come from, which is still hard for us to answer. Maybe
tension of the branes in string theory is a nice candidate for the
positive cosmological constant.

\vspace{.5cm}

\noindent {\bf Acknowledgments}

We would like to thank M. Li and H. Tye for useful comments. This
work was supported by a grant from NSFC, a grant from China
Postdoctoral Science Foundation and a grant from K. C. Wang
Postdoctoral Foundation.

\end{document}